\documentclass[conference,compsoc]{IEEEtran}

\ifCLASSOPTIONcompsoc
  \usepackage[nocompress]{cite}
\else
  \usepackage{cite}
\fi

\ifCLASSINFOpdf
\else
\fi
\usepackage{graphicx}
\usepackage{amsmath}
\usepackage{booktabs}

\begin{document}
\title{Machine Learning Based Framework for Estimation of Data Center Power \\ Using Acoustic Side Channel}

\author{\IEEEauthorblockN{Mohsen Karimi}
\IEEEauthorblockA{University of California, Riverside\\
Riverside, California, USA\\
mkari007@ucr.edu}
\and
\IEEEauthorblockN{Fahimeh Arab}
\IEEEauthorblockA{University of California, Riverside\\
Riverside, California, USA\\
farab002@ucr.edu}
}


% compsoc conferences compared to traditional conferences), use this
%Telephone: (800) 555--1212, Fax: (888) 555--1212}


\maketitle
\thispagestyle{plain}
\pagestyle{plain}
\begin{abstract}
Data centers are high power consumers and the energy consumption of data centers keeps on rising in spite of all the efforts for increasing the energy efficiency. The need for energy-awareness in data centers makes the use of power modeling and estimation to be still a big challenge due to huge amount of uncertainty in this area. In this paper, a machine learning based method is proposed to approximately estimate the amount of power consumption by using acoustic side channel caused by fan in the fan-based cooling system in the server room. For doing so, frequency components of the acoustic signal, recorded by a microphone in the server room, is extracted, pre-processed, and fed to a Multi-Layer Neural-Network as an estimator. The proposed method performed well to estimate the power consumption, having more than 85 percent accuracy.
\end{abstract}
\begin{IEEEkeywords}
Power Estimation, Data Center, Classification, Neural Networks, Acoustic Side Channel
\end{IEEEkeywords}


% \fi
\IEEEpeerreviewmaketitle

\section{Introduction}\label{IntroductionSection}

Energy consumed by data centers have been increased more than 55\% from 2005 to 2010 and even with current advancements in energy efficiency of the devices and energy reduction techniques, it is expected to increase in the following years\cite{USEnergy16}. This has encouraged the further development of techniques to reduce the energy consumption and the environmental footprint of data centers \cite{PowerModeling16}. These techniques require data centers to be energy-aware, that is, they must be able to measure and predict their energy consumption. 

A common approach toward energy-awareness is through power models, which allow estimating the power consumption by means of indirect evidences (such as resource usage) \cite{trans14} or through effect of the server's power consumption on data center room, which are called Side-Channels. These side-channels can be any environmental changes that the power consumption cause due to increase or decrease of energy demand of the servers which can be the temperature, sound noise, or air disturbance of the room, to name but a few.

Power estimation differs from power measurement, which measures the actual power consumption by means of special hardware devices. Power models or estimations are especially helpful when direct measurement is not possible (e.g. to assess power of individual software components such as processes or virtual machines) or it is expensive (e.g. to assess power at low granularities). Since some of these estimations are done by using effects of server's power consumption on server room, they can be done by other tenants on a multi-tenant server. This information can be used by malicious tenants to find other server's vulnerable time (i.e. power peaks) to attack these servers\cite{Islam17}. Therefore, it is important to know the vulnerable time of the server by using power estimations based on side channels like thermal and acoustic side channels. 

Machine learning has been used and showed promising performance in many classification applications from computer networks \cite{bgp_bigdata, ns3_qos} to behavior classification in humans and animals \cite{abdoli1, abdoli2}, which used a novel hybrid classification algorithm which attempts to conduct the classification task at hand through a combination of shape and feature classification methods that has resulted in improved classification accuracy. This is in contrast to traditional classification methods, that merely rely on either shape classification or feature classification alone.

In this work, we aim to present a power estimation system based on acoustic side channel using Neural Network as the estimator. Generally, side channel based techniques try to estimate a phenomenon (here power consumption) using effects of that phenomenon on the environment. The main purpose of this method is to classify the state of the server to four classes of Lowest-Power-Consuming, Low-Power-Consuming, High-Power-Consuming, and Highest-Power-Consuming by using acoustic side channel. The acoustic data we used in this work is recorded in a server room located at University of California Riverside\cite{Islam17}.

The rest of the paper is structured as follows: In Section \ref{DataSection}, structure of the data-sets and its features are presented. Some the pre-processing techniques used to extract features, and specification of the proposed method are illustrated in \ref{MethodSection}. Results of the network is presented in Section \ref{ResultsSection}, and finally conclusion and some suggestions for future works are presented in the Section \ref{ConclusionSection}.

\section{Data Preparation}\label{DataSection}

Although liquid cooling systems, which circulate a liquid through a heat sink attached to the processor, are also used in some computing centers, fan based air cooling systems are the most commonly cooling systems used in the multi-tenant data centers. In a server room, different kinds of noises such as air conditioning (AC) noises, server fan noises, hard drives, and electrical components such as capacitors and transformers exist. However, the most dominant noise sources are cooling fan noises from the servers and fan noises produced by air conditioning systems.

The rotating blades in a server's cooling fans create pulsating variations in air pressure levels which is shown in Figure \ref{fanfig}. As it is shown in Figure \ref{fanfig}, the rotating fan generates a high-pitched noise. The frequency component of the noise depends on the fan speed. The relationship between the noise major tone frequency and fan speed in RPM (revolutions per minute) is governed by: $Frequency (Hz) = 1/60*Fan\:Speed (in\:RPM) * Number\:of\:blades$ \cite{Islam17,dell11}. Since most of the fans that are used in data centers have 5 to 7 blades and the speed of the fans are between 2000 to 6000 RPM, the frequency of the noise generated by server fans would be between 166 to 700 Hertz. From the experiments done by \cite{Islam17}, the noise generated by air conditioning facilities has the frequency components lower than 200 Hz.

In this work, the data is gathered from recorded noise of a server used in \cite{Islam17} which is also available on \cite{IslamWeb}. The sound is recorded in two different environments. In the first experiment, noise was recorded with a microphone from the running servers located in a quiet laboratory. In the second experiment, noise was recorded from the running servers located in a real multi-tenant data center which consists of multiple tenant's servers and AC facilities all making noise.
Followings are the specifications for each of these above mentioned experiments:

\begin{figure}
\centering
\includegraphics[width=0.36\textwidth]{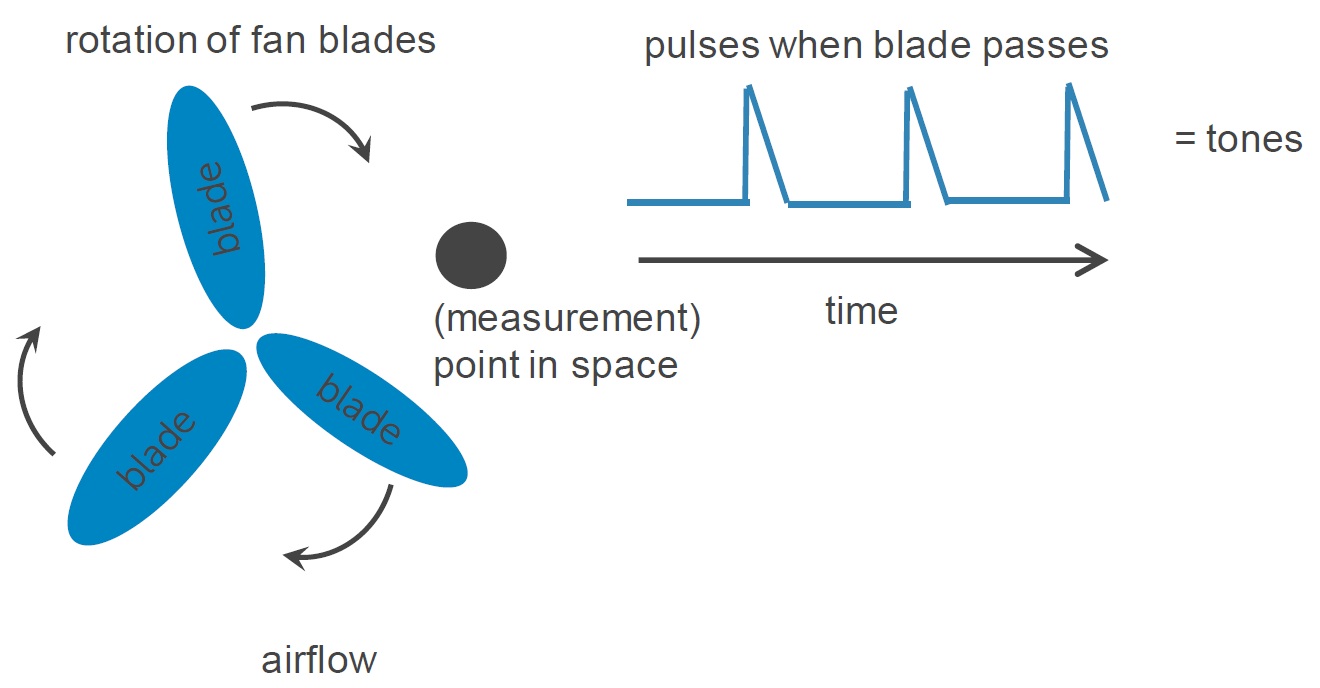}
\caption{Fan airborne noise mechanism\cite{dell11}}
\label{fanfig}
\end{figure}

\begin{enumerate}
\item Noise recorded in the lab environment
    \begin{itemize}
        \item Sound file: noise\_recording.wav
        \begin{itemize}
            \item Total duration:  5:55:00
            \item Sampling Freq: 16kHz
        \end{itemize} 
        \item Power data: power\_data.csv
        \begin{itemize}
            \item Sampling rate: 1 sample/20s
        \end{itemize}     
    \end{itemize}
    
\item Noise recorded in the multi-tenant data center
    \begin{itemize}
        \item Sound file: server\_in\_data\_center.wav
        \begin{itemize}
            \item Total duration:  5:55:00
            \item Sampling Freq: 16kHz
        \end{itemize} 
        \item Power data: \\power\_data\_server\_in\_data\_center.cs
        \begin{itemize}
            \item Sampling rate: 1 sample/20s
        \end{itemize}     
    \end{itemize}
\end{enumerate}

\begin{figure}
\centering
\includegraphics[width=0.4\textwidth]{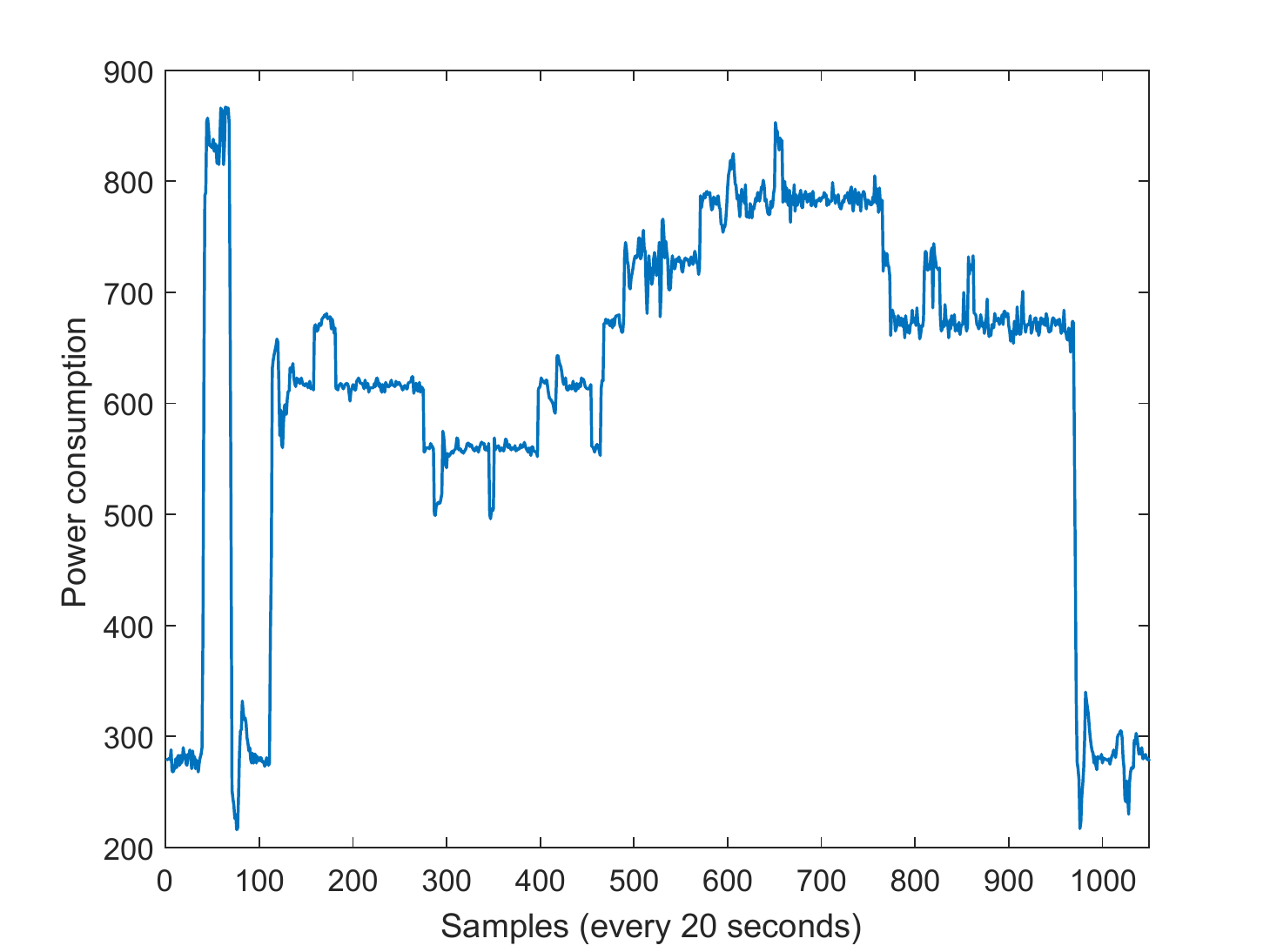}
\caption{Power consumption samples with AC noise}
\label{powerfig}
\end{figure}

The server power consumption samples for every 20 second with the presence of AC noise is also shown in Figure \ref{powerfig}

\section{Proposed Methodology}\label{MethodSection}
In this work, we use frequency features of each segment of signal and a Neural Network classifier to assign the state of the server into four power consumption classes. The features are extracted from the power spectral density of the fan noise sound captured in the room.
\subsection{Feature Extraction}\label{FeatureSection}
In every classification problem, one of the most important steps is extracting features based on data behavior in a way that the classes can be distinguished from each other properly. In this work, two approaches are introduced depending on frequency behavior of fan noise data. As discussed from the previous sections, the frequency of fan noise is in the range of 166 to 600 Hz and the frequency of AC noise is less than 200 Hz. Therefore, by using a band-pass filter, we can achieve the desired frequency components corresponding to the servers fan noise. One of the methods which can be used for extracting frequency components of a digital signal is using Discrete Fourier Transform which can be calculated from equation bellow \cite{fft05}.
\begin{equation}
\label{dft}
  X_k =  \sum_{n=0}^{N-1} x_n e^{-i2\pi kn/N} \qquad
\end{equation}

Where $N$ is the number of samples which here is $20 \:(Seconds) * 16000 \:(Hz) = 320000$, and $X_k$ is $k$th frequency component of the signal.  Since the desired frequency components are in 166 to 600 Hz interval, the desired components of Fourier transform would be for $k= 3320 \: to\: 14000$. Power of $k$th frequency component can be calculated from the following equation:

\begin{equation}
\label{powerdft}
    |X_k| = \sqrt{\operatorname{Re}(X_k)^2 + \operatorname{Im}(X_k)^2}
\end{equation}

In the first approach, all the frequency components calculated from the Equation \ref{powerdft} can be used as the inputs of the classifier. \par
From the total time duration ($T$) and sampling frequency ($f_s$) frequency resolution (i.e. frequency distance between two consecutive samples) of the Fourier transform would be $\Delta f_{min} = \frac{f_s}{N} = \frac{16000}{320000} = 0.05Hz$ which is more than enough for our purpose. Since the speed of the fan does not have a very large granularity, having a very high resolution in frequency makes our classifier more complex and less accurate by feeding some extra non-relevant information as its input. \par

Thus, we divided the desired frequency span into $15Hz$ sub-regions and then used the component with maximum power component as the representative of the whole sub-region. By doing so, the number of inputs to the classifier would decrease by the factor of $\frac{1}{300}$. we used these reduced number of selected frequency components as the inputs to the classifier in the second approach.\par

It should be noted that the frequency range for sub-regions, i.e. 15Hz, is chosen heuristically and based on trial and error to achieve best performance from the classifier.

\begin{figure}
\centering
\includegraphics[width=0.4\textwidth]{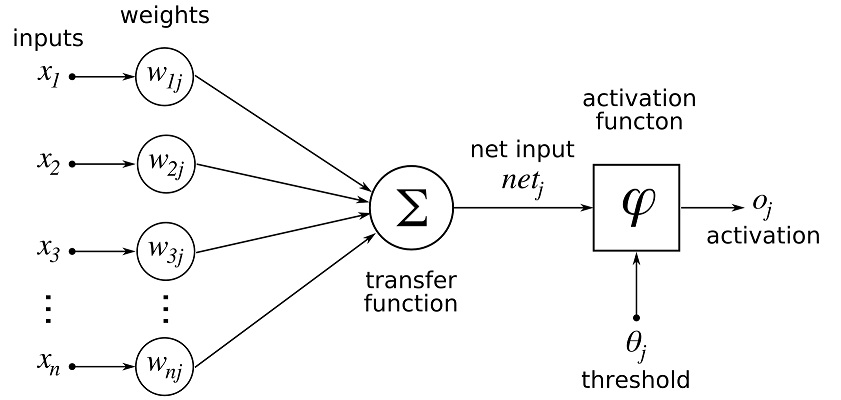}
\caption{A Neuron scheme in an artificial neural network\cite{McCulloch1943}}
\label{neuronfig}
\end{figure}

\begin{figure}
\centering
\includegraphics[width=0.45\textwidth]{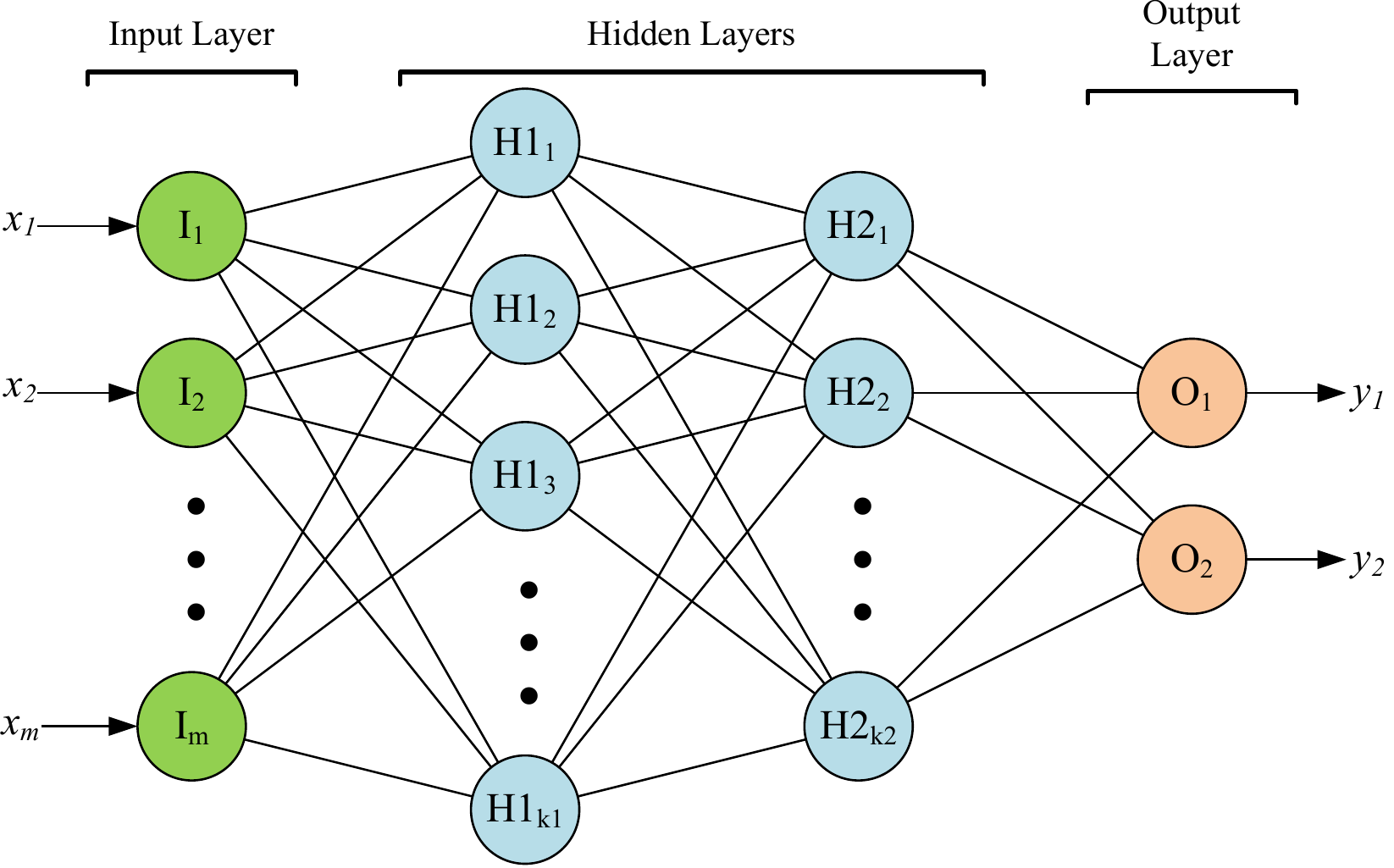}
\caption{Neural Network scheme used in this work}
\label{networkfig}
\end{figure}

\subsection{Neural Network}\label{NetworkSection}
A Back Propagation Neural Network is a multi-layer combination of  components called neurons. The main process on signal is the forward pass, then the error back propagation. Neural Networks have been widely used in image processing, function approximation, data mining and so on\cite{Staszewski1997}. \par
The whole network is consists of an input layer, an output layer, and some layers in between which are called hidden layers. The direction of the forward pass is from input layer, hidden layer to output layer. The neural state of each layer can only influence the next layer neural state. Since we set the expected output corresponding to the input, if the BP neural network output cannot get the expected output, it will turn to back propagation and adjust the network structure through gradient descent method in the process of back propagation to minimize the error. By repeating this process, we can obtain a model to express the relationship between input and output\cite{Zhou15}. \par
The question of how many hidden layers and how many hidden nodes should always come up in any classification task using neural networks. Until today there has been no exact solution. A method of shedding some light to this question is presented in \cite{Stathakis09} which a near‐optimal solution is discovered after searching with a genetic algorithm. In this work, number of hidden layers are chosen 2 and number of neurons in each hidden layer is chosen based on Equation \ref{Hidden_Layer_1} and Equation \ref{Hidden_Layer_2} which are the most optimum numbers according to\cite{Stathakis09} which is also used in \cite{bgp_bigdata}:

\begin{equation}
\label{Hidden_Layer_1}
k1=\sqrt{m*(p+2)}+2\sqrt{\frac{p}{p+2}}
\end{equation}

\begin{equation}\label{Hidden_Layer_2}
k2=p\sqrt{\frac{m}{p+2}}
\end{equation}

Where p, m, k1, and k2 are number of outputs, number of inputs, number of neurons in the first hidden layer, and the number of neurons in the second hidden layer respectively. \par

The overall scheme of each neuron is shown in Figure \ref{neuronfig}. Where $(x_1, ... , x_n)$ is the signal vector that gets multiplied with the weights $\left( w_{1j}, w_{2j}, ..., w_{nj} \right)$. This is followed by accumulation (i.e. summation + addition of bias $b$). Finally, an activation function $f$ is applied to this sum.

Note that the weights $\left( w_1, w_2, \cdots, w_n \right)$ and the biases $b$ transform the input signal linearly. The activation, on the other hand, transforms the signal non-linearly and it is this non-linearity that allows us to learn arbitrarily complex transformations between the input and the output\cite{McCulloch1943}.

In this work he activation function used to calculate each layer's output from the summation of weighted inputs is based on tansig function which is the Hyperbolic Tangent Sigmoid. This activation function has output range from -1 to +1 that is very suitable in cases where there are only two classes. The formula for sigmoid function is given in Equation \ref{tansig_eq}

\begin{equation}\label{tansig_eq}
tansig(x)=\frac{2}{1+e^{-2x}}-1
\end{equation}

After last stage of the network, a simple threshold function is used to achieve -1 or 1 numbers. This stage is implemented separately outside of the network. The threshold function used in the paper is shown in Equation \ref{threshold}
\begin{equation}\label{threshold}
f(x)= \begin{cases}
  -1 & x<0\\
  1 & x\geq0
\end{cases}
\end{equation}

The power consumption of the server is divided into four classes. For doing so, we divided the range of power consumption (i.e. minimum to maximum power consumption) into four equal intervals and assigned all the power samples corresponding to each interval to a class. For classification purpose we assign each class to a different sets of -1 and 1: (1, 1) to class 1, (-1, 1) to class 2, (1, -1) to class 3, and (-1, -1) to class 4. This assignment is done to improve the performance of the classifier based on the output of transfer function chosen in the network. Thus, the number of output neurons would be 2. The network scheme is also shown in Figure \ref{networkfig}.

Other settings of the network are set using trial and error to achieve the best performance. In summary, Table \ref{networktable} shows a summary of the neural network specifications used in this paper.

\begin{table}
\centering
\caption{Overview of Neural network specification}
\label{networktable}
\begin{tabular}{|l|c|}
\hline
\multicolumn{2}{|c|}{\textbf{Neural Network Specification}} \\ \hline
Input Layers & m Neurons \\ \hline
Output Layer & 2 Neuron \\ \hline
First Hidden Layer & k1 Neurons \\ \hline
Second Hidden Layer & k2 Neurons \\ \hline
Activation Function & Sigmoid \\ \hline
Maximum Number of Epochs & 1000 \\ \hline
Training Function & Gradient Descent \\ \hline
Simulation Framework & Matlab \\ \hline
Goal Error & 1e-4 \\ \hline
\end{tabular}
\end{table}

It should be noted that the initial weights of the network are chosen randomly and the network is trained based on the gradient descent algorithm, the trained network is not exactly the same when running the same network multiple times with the same inputs and outputs in the training process.

\begin{figure}
\centering
\includegraphics[width=0.4\textwidth]{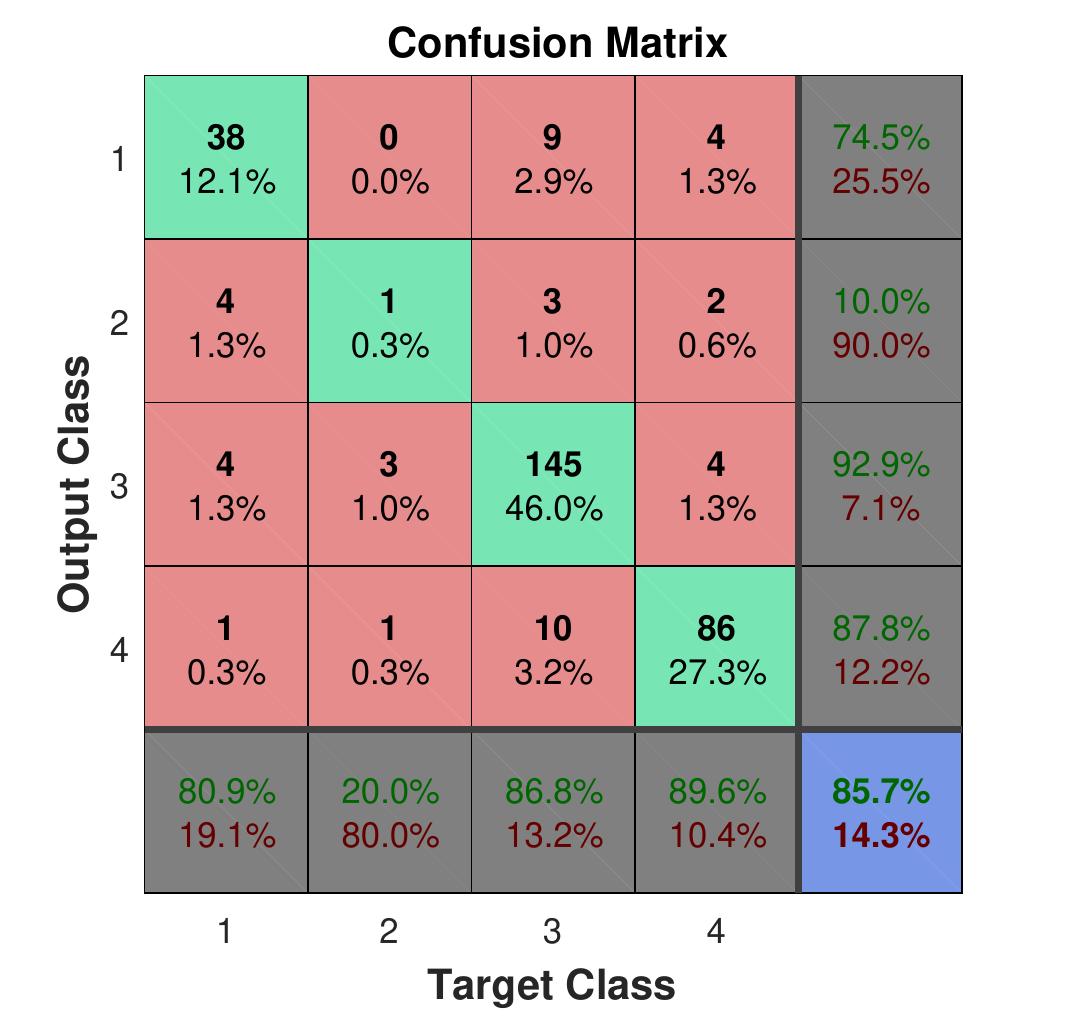}
\caption{Confusion matrix of tested data on second approach when AC noise is considered}
\label{confusionfig}
\end{figure}

\section{Results Analysis}\label{ResultsSection}
\begin{figure*}
\includegraphics[width=0.7\textwidth]{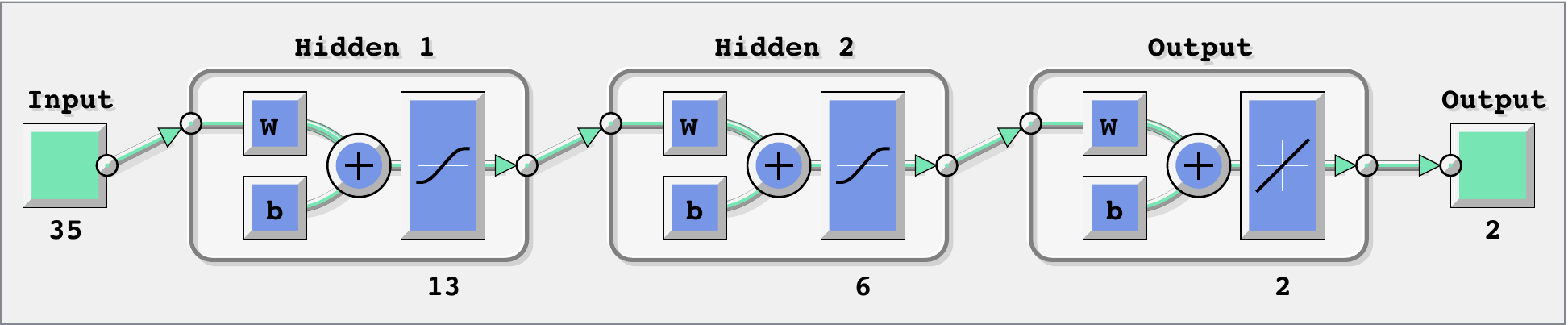}
\centering
\caption{Neural Network layers for reduced frequency components method}
\label{networkviewfig}
\end{figure*}

Two different method, i.e. using frequency components versus using reduced frequency components, with two different types of data, i.e. noise recorded without AC noise versus noise recorded with AC noise were fed to the neural network classifier illustrated in Section \ref{NetworkSection}. The Scheme of the Network is show in figure \ref{networkviewfig}. 70\% of data has been randomly chosen for training step, and the reset 30\% has been used for testing step to evaluate the performance of the trained network.\par  
The accuracy of different methods with different noise data is shown in Table \ref{ResultsTable}. Although we could not achieve an outstanding performance in the second method when AC noise is considered, there is a significant improvement when using second method compared to using the first method. \par
One of the best tools to see the performance of a classifier is the confusion matrix. A confusion matrix is a table that is often used to describe the performance of a classification model on a set of test data for which the true values are known. Confusion matrix of the tested data for second method when AC noise is considered is shown Figure \ref{confusionfig}. 

\begin{figure}
\centering
\includegraphics[width=0.4\textwidth]{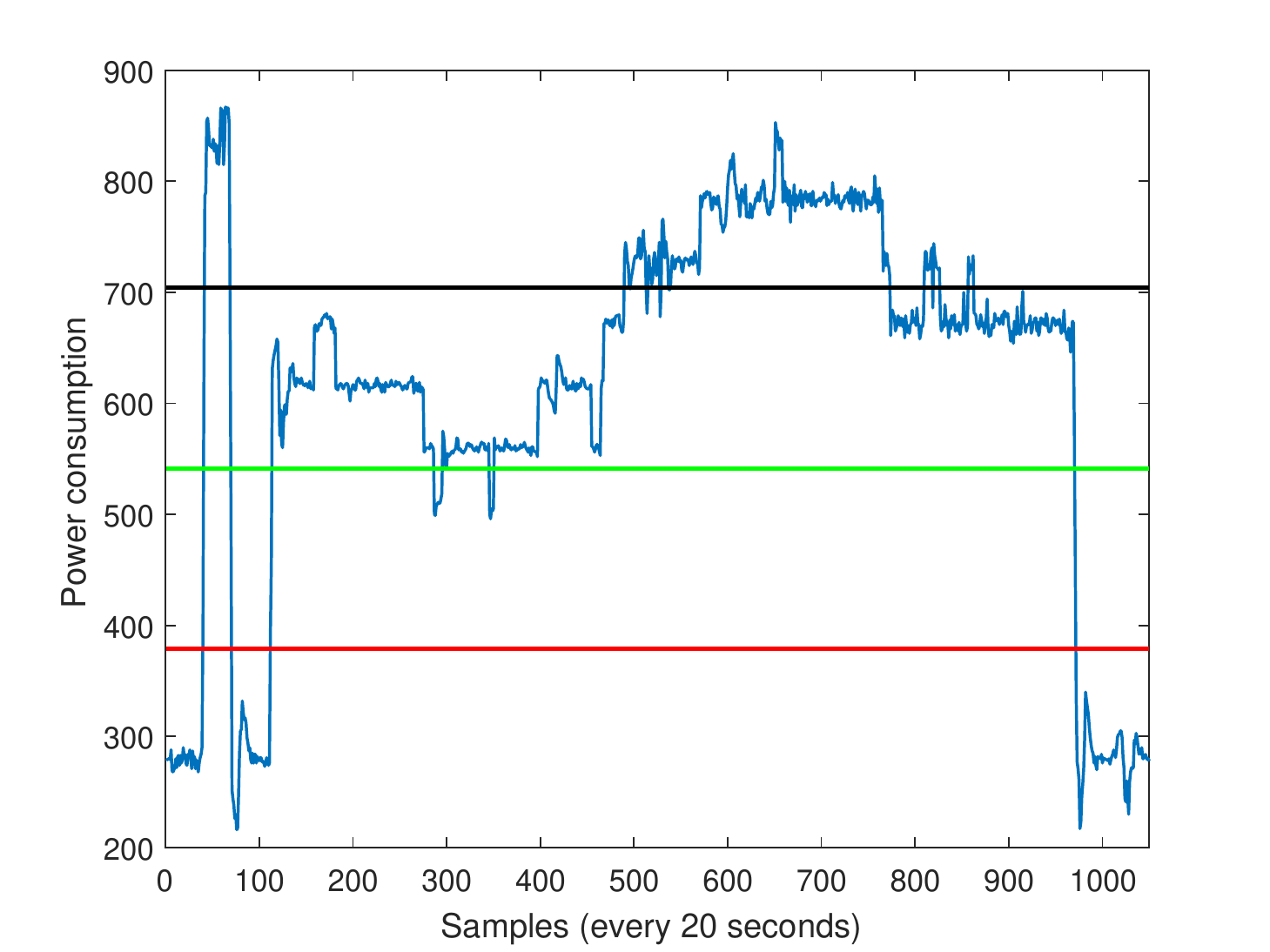}
\caption{Different classes of power consumption samples with AC noises}
\label{powerclassfig}
\end{figure}

\begin{table}
\centering

  \caption{Accuracy of the classifier}
  \label{ResultsTable}
  \begin{tabular}{l|c}
    \toprule
    \multicolumn{1}{c|}{Method} & Accuracy\\
    \midrule
    {First approach on sound without AC noise} & 93\% \\
    {First approach on sound with AC noise}& 56\% \\
    {Second approach on sound without AC noise}& 97\%\\
    {Second approach on sound with AC noise}& 85\%\\
    \bottomrule
  \end{tabular}
\end{table}

As we can see from Figure \ref{confusionfig}, worst performance of the classifier is achieved for the second class. If we take a look at the four classes in the power samples which is shown in figure \ref{powerclassfig}, we can see one of the reasons that may cause this behavior. As shown in Figure \ref{powerclassfig}, second class samples (samples between red and green horizontal lines) are comparatively fewer than other classes. This phenomenon, would cause the classifier to have few samples from this class and therefore perform worse in testing step.\par
Furthermore, since the part of the data which was used as training data was chosen randomly, the amount of data from each class that was used for the training step may be very different and in some cases data of one class may not be used in the training step.

\section{Conclusion and Future Work}\label{ConclusionSection}

In this paper we proposed a new method to build a classifier to classify power consumption of a multi-tenant server. To classify data we used frequency features based on acoustic side channels from a multi-tenant server. The data of two different scenarios with and without AC noise were used to train and test the classifier. The features were fed to an Multi-Layer Neural Network to build the classifier. At the end, the experimental results of two method were discussed and compared.\par

One of the most important part of every classification algorithm is the feature selection and feature extraction parts. In this paper, we used frequency components of the sound noise due to relationship between fan speed and frequency of the noise it generates. However, adding more features to the feature vector would possibly result in better performance of the trained classifier. Another direction for future works could be using more advanced network like convolutional neural networks to extract features from dataset by the network itself. This could result in extracting features which may be more effective for these kinds of anomalies.\par

As we see from the Section \ref{ResultsSection}, one of the main reasons that caused depression in our performance was the not-optimum class definition. A better definition of classes, in the way that lead to the number of samples in each class to be equal, would result in better performance in the trained classifier.



%\bibliographystyle{IEEEtran}
\bibliographystyle{IEEEtran}
\bibliography{main.bib}

\end{document}